\documentclass[onecolumn,showpacs,preprintnumbers,amsmath,amssymb,eqsecnum]{revtex4}
\usepackage{dcolumn}
\usepackage{bm}
\usepackage[dvips]{graphicx}
\usepackage{wrapfig}
\usepackage{psfrag}
\begin{document}

\def\sh{\mathop{\rm sh}\nolimits}
\def\ch{\mathop{\rm ch}\nolimits}
\def\var{\mathop{\rm var}}\def\exp{\mathop{\rm exp}\nolimits}
\def\Re{\mathop{\rm Re}\nolimits}
\def\Sp{\mathop{\rm Sp}\nolimits}
\def\kp{\mathop{\text{\ae}}\nolimits}
\def\bk{{\bf {k}}}
\def\bp{{\bf {p}}}
\def\bq{{\bf {q}}}
\def\lra{\mathop{\longrightarrow}}
\def\Const{\mathop{\rm Const}\nolimits}
\def\sh{\mathop{\rm sh}\nolimits}
\def\ch{\mathop{\rm ch}\nolimits}
\def\var{\mathop{\rm var}}
\def\mK{\mathop{{\mathfrak {K}}}\nolimits}
\def\mR{\mathop{{\mathfrak {R}}}\nolimits}
\def\mv{\mathop{{\mathfrak {v}}}\nolimits}
\def\mV{\mathop{{\mathfrak {V}}}\nolimits}
\def\mD{\mathop{{\mathfrak {D}}}\nolimits}
\def\mN{\mathop{{\mathfrak {N}}}\nolimits}
\def\mS{\mathop{{\mathfrak {S}}}\nolimits}

\newcommand\ve[1]{{\mathbf{#1}}}

\def\Re{\mbox {Re}}
\newcommand{\Z}{\mathbb{Z}}
\newcommand{\R}{\mathbb{R}}
\def\mK{\mathop{{\mathfrak {K}}}\nolimits}
\def\mL{\mathop{{\mathfrak {L}}}\nolimits}
\def\mk{\mathop{{\mathfrak {k}}}\nolimits}
\def\mR{\mathop{{\mathfrak {R}}}\nolimits}
\def\mv{\mathop{{\mathfrak {v}}}\nolimits}
\def\mV{\mathop{{\mathfrak {V}}}\nolimits}
\def\mD{\mathop{{\mathfrak {D}}}\nolimits}
\def\mN{\mathop{{\mathfrak {N}}}\nolimits}
\def\ml{\mathop{{\mathfrak {l}}}\nolimits}
\def\mf{\mathop{{\mathfrak {f}}}\nolimits}
\def\me{\mathop{{\mathfrak {e}}}\nolimits}
\newcommand{\ccm}{{\cal M}}
\newcommand{\cE}{{\cal E}}
\newcommand{\cV}{{\cal V}}
\newcommand{\cI}{{\cal I}}
\newcommand{\cR}{{\cal R}}
\newcommand{\cK}{{\cal K}}
\newcommand{\cH}{{\cal H}}
\newcommand{\cW}{{\cal W}}
\newcommand{\cS}{{\cal S}}
\newcommand{\cT}{{\cal T}}
\newcommand{\ce}{{\cal e}}

\def\br{\mathop{{\bf {r}}}\nolimits}
\def\bS{\mathop{{\bf {S}}}\nolimits}
\def\bA{\mathop{{\bf {A}}}\nolimits}
\def\bJ{\mathop{{\bf {J}}}\nolimits}
\def\bn{\mathop{{\bf {n}}}\nolimits}
\def\bg{\mathop{{\bf {g}}}\nolimits}
\def\bv{\mathop{{\bf {v}}}\nolimits}
\def\be{\mathop{{\bf {e}}}\nolimits}
\def\bp{\mathop{{\bf {p}}}\nolimits}
\def\bz{\mathop{{\bf {z}}}\nolimits}
\def\bbf{\mathop{{\bf {f}}}\nolimits}
\def\bb{\mathop{{\bf {b}}}\nolimits}
\def\ba{\mathop{{\bf {a}}}\nolimits}
\def\bx{\mathop{{\bf {x}}}\nolimits}
\def\by{\mathop{{\bf {y}}}\nolimits}
\def\br{\mathop{{\bf {r}}}\nolimits}
\def\bs{\mathop{{\bf {s}}}\nolimits}
\def\bH{\mathop{{\bf {H}}}\nolimits}
\def\bk{\mathop{{\bf {k}}}\nolimits}
\def\be{\mathop{{\bf {e}}}\nolimits}
\def\bnul{\mathop{{\bf {0}}}\nolimits}
\def\bq{{\bf {q}}}

\newcommand{\oV}{\overline{V}}
\newcommand{\vkp}{\varkappa}
\newcommand{\os}{\overline{s}}
\newcommand{\opsi}{\overline{\psi}}
\newcommand{\ov}{\overline{v}}
\newcommand{\oW}{\overline{W}}
\newcommand{\oPhi}{\overline{\Phi}}

\def\mI{\mathop{{\mathfrak {I}}}\nolimits}
\def\mA{\mathop{{\mathfrak {A}}}\nolimits}

\def\diag{\mathop{\rm diag}\nolimits}
\def\st{\mathop{\rm st}\nolimits}
\def\tr{\mathop{\rm tr}\nolimits}
\def\sign{\mathop{\rm sign}\nolimits}
\def\d{\mathop{\rm d}\nolimits}
\def\const{\mathop{\rm const}\nolimits}
\def\O{\mathop{\rm O}\nolimits}
\def\Spin{\mathop{\rm Spin}\nolimits}
\def\exp{\mathop{\rm exp}\nolimits}
\def\SU{\mathop{\rm SU}\nolimits}
\def\mU{\mathop{{\mathfrak {U}}}\nolimits}
\newcommand{\cU}{{\cal U}}
\newcommand{\cD}{{\cal D}}

\def\mC{\mathop{{\mathfrak {C}}}\nolimits}
\def\mI{\mathop{{\mathfrak {I}}}\nolimits}
\def\mA{\mathop{{\mathfrak {A}}}\nolimits}
\def\mU{\mathop{{\mathfrak {U}}}\nolimits}

\def\st{\mathop{\rm st}\nolimits}
\def\tr{\mathop{\rm tr}\nolimits}
\def\sign{\mathop{\rm sign}\nolimits}
\def\d{\mathop{\rm d}\nolimits}
\def\const{\mathop{\rm const}\nolimits}
\def\O{\mathop{\rm O}\nolimits}
\def\Spin{\mathop{\rm Spin}\nolimits}
\def\exp{\mathop{\rm exp}\nolimits}

\title{A note on the vacuum structure of lattice Euclidean quantum gravity: “birth” of
macroscopic space-time and PT-symmetry breaking}

\author {S.N. Vergeles\vspace*{4mm}\footnote{{e-mail:vergeles@itp.ac.ru}}}

\affiliation{Landau Institute for Theoretical Physics,
Russian Academy of Sciences,
Chernogolovka, Moscow region, 142432 Russia \linebreak
and   \linebreak
Moscow Institute of Physics and Technology, Department
of Theoretical Physics, Dolgoprudnyj, Moskow region,
141707 Russia}

\begin{abstract} It is shown that the ground state or vacuum of the lattice quantum gravity
is significantly different from the ground states of the well-known vacua in QED, QCD, et cetera.
In the case of the lattice quantum gravity, the long-wavelength scale vacuum structure is
similar to that in QED, moreover the quantum fluctuations of gravitational degrees of freedom are very reduced in comparison with the situation in QED. But the small scale (of the order of the lattice scale) vacuum structure in gravity is significantly different from that in the long-wavelength scales: the fluctuation values of geometrical degrees of freedom (tetrads) are commensurable with  theirs most probable values. It is also shown that the macroscopic Universe can exist only in the presence of fermion fields. In this case, spontaneous breaking of the PT symmetry occurs.

\end{abstract}


\maketitle

\section{Introduction}

The work shows that the system of gravity coupled with Dirac or Weyl fermions breaks PT-invariance.
The lattice version of the regularization of the theory of gravity is used. Vacuum fluctuations are studied both in the lattice theory of pure gravity and in the case of lattice gravity coupled with fermions.
It is shown that in the pure theory of gravitation the birth of macroscopic space-time is impossible.
Such a possibility appears when fermionic fields are included in the theory.

Let's outline the model of lattice gravity which is studied here.
A detailed description of the model and some of its properties arwe given in \cite{vergeles2015wilson,vergeles2016instanton,vergeles2006one,vergeles2008doubling,vergeles2017note,vergeles2017fermion}.

In this section, consideration is carried out for the Euclidean signature, since it is difficult to distinguish the temporal direction on a disordered lattice.

Denote by $\gamma^a$ $4\times4$ Hermitian Dirac matrices, so that
\begin{gather}
\gamma^a\gamma^b+\gamma^b\gamma^a=2\delta^{ab}, \quad \sigma^{ab}\equiv\frac{1}{4}[\gamma^a,\gamma^b], \quad a=1,2,3,4, \quad
\gamma^5\equiv\gamma^1\gamma^2\gamma^3\gamma^4=(\gamma^5)^{\dag}.
\label{Dirac_Algebra}
\end{gather}

Consider the orientable abstract 4-dimensional simplicial complex  $\mK$.
Suppose that any of its 4-simplexes belongs to such a finite (or infinite) sub-complex ${\mK}'\in\mK$  which has a geometric realization in  $\R^4$ topologically equivalent to a disk  without cavities.
The vertices are designated as
$a_{\cV}$, the indices ${\cV}=1,2,\dots,\,{\mN}^{(0)}\rightarrow\infty$ and ${\cW}$ enumerate the vertices
and 4-simplices, correspondingly; the pairs of indices $({\cV}_1{\cV}_2)=({\cV}_2{\cV}_1)$ enumerate 1-simplexes $a_{{\cV}_1}a_{{\cV}_2}\in\mK$
regardless of their orientation. It is necessary to use
the local enumeration of the vertices $a_{\cV}$ attached to a given
4-simplex: the all five vertices of a 4-simplex with index ${\cW}$
are enumerated as $a_{{\cV}_{({\cW})i}}$, $i=1,2,3,4,5$. The later notations with extra low index  $({\cW})$
indicate that the corresponding quantities belong to the
4-simplex with index ${\cW}$. Of course, these same quantities  also belong to another 4-simplex with index  ${\cW}'$, and 4-simplexes with indices  ${\cW}$ and  ${\cW}'$ must be adjacent.
The Levi-Civita symbol with in pairs
different indexes
$\varepsilon_{{\cV}_{({\cW})1}{\cV}_{({\cW})2}{\cV}_{({\cW})3}{\cV}_{({\cW})4}{\cV}_{({\cW})5}}=\pm 1$ depending on
whether the order of vertices
$s^4_{\cW}=a_{{\cV}_{({\cW})1}}a_{{\cV}_{({\cW})2}}a_{{\cV}_{({\cW})3}}a_{{\cV}_{({\cW})4}}a_{{\cV}_{({\cW})5}}$ defines the
positive or negative orientation of 4-simplex $s^4_{\cW}$.
An element of the compact group $\Spin(4)$ and an element of the Clifford algebra
\begin{gather}
\Omega_{{\cV}_1{\cV}_2}=\Omega^{-1}_{{\cV}_2{\cV}_1}=
\exp\left(\omega_{{\cV}_1{\cV}_2}\right)
\in\Spin(4), \quad
\omega_{{\cV}_1{\cV}_2}\equiv\frac{1}{2}\sigma^{ab}
\omega^{ab}_{{\cV}_1{\cV}_2}=-\omega_{{\cV}_2{\cV}_1},
\nonumber \\
\hat{e}_{{\cV}_1{\cV}_2}\equiv
-\Omega_{{\cV}_1{\cV}_2}\hat{e}_{{\cV}_2{\cV}_1}\Omega_{{\cV}_1{\cV}_2}^{-1}
\equiv e^a_{{\cV}_1{\cV}_2}\gamma^a,  \quad
-\infty<e^a_{{\cV}_1{\cV}_2}<+\infty,
\label{Variables_Grav}
\end{gather}
are assigned for each oriented 1-simplex $a_{{\cV}_1}a_{{\cV}_2}$.

The conjecture is that the set of variables $\{\Omega,\,\hat{e}\}$  is an independent set of dynamic variables.
Fermionic degrees of freedom (Dirac spinors) are assigned to each vertex of the complex:
\begin{gather}
\Psi^{\dag}_{\cV}, \quad \Psi_{\cV}.
\label{Variables_Ferm}
\end{gather}
The set of variables $\{\Psi^{\dag},\,\Psi\}$ is also a set of mutually independent variables, and the spinors $\Psi^{\dag}_{\cV}$ and $\Psi_{\cV}$ are in mutual involution (or anti-involution) relative to Hermitian conjugation operation.
\begin{gather}
\mA=\mA_g+\mA_{\Psi},
\label{Latt_Action}
\end{gather}
\begin{gather}
\mA_g=\frac{1}{5\cdot
24\cdot2\cdot l_P^2}\sum_{\cW}\sum_{\sigma}
\varepsilon_{\sigma({\cV}_{({\cW})1})\sigma({\cV}_{({\cW})2})
\sigma({\cV}_{({\cW})3})\sigma({\cV}_{({\cW})4})\sigma({\cV}_{({\cW})5})}
\nonumber \\
\times\tr\gamma^5\bigg\{
\Omega_{\sigma({\cV}_{({\cW})5})\sigma({\cV}_{({\cW})1})}
\Omega_{\sigma({\cV}_{({\cW})1})\sigma({\cV}_{({\cW})2})}\Omega_{\sigma({\cV}_{({\cW})2})\sigma({\cV}_{({\cW})5})}
\hat{e}_{\sigma({\cV}_{({\cW})5})\sigma({\cV}_{({\cW})3})}
\hat{e}_{\sigma({\cV}_{({\cW})5})\sigma({\cV}_{({\cW})4})}\bigg\},
\label{Latt_Action_Grav}
\end{gather}
\begin{gather}
\mA_{\Psi}=-\frac{1}{5\cdot24^2}\sum_{\cW}\sum_{\sigma}
\varepsilon_{\sigma({\cV}_{({\cW})1})\sigma({\cV}_{({\cW})2})
\sigma({\cV}_{({\cW})3})\sigma({\cV}_{({\cW})4})\sigma({\cV}_{({\cW})5})}
\nonumber \\
\times\tr\gamma^5\bigg\{ \hat{\Theta}_{\sigma({\cV}_{({\cW})5})\sigma({\cV}_{({\cW})1})}
\hat{e}_{\sigma({\cV}_{({\cW})5})\sigma({\cV}_{({\cW})2})}
\hat{e}_{\sigma({\cV}_{({\cW})5})\sigma({\cV}_{({\cW})3})}
\hat{e}_{\sigma({\cV}_{({\cW})5})\sigma({\cV}_{({\cW})4})}\bigg\}
\label{Latt_Action_Ferm}
\end{gather}
\begin{gather}
\hat{\Theta}_{{\cV}_1{\cV}_2}\equiv
\frac{i}{2}\gamma^a\left(\Psi^{\dag}_{{\cV}_1}\gamma^a
\Omega_{{\cV}_1{\cV_2}}\Psi_{{\cV}_2}-\Psi^{\dag}_{{\cV}_2}\Omega_{{\cV}_2{\cV}_1}\gamma^a\Psi_{{\cV}_1}\right)
\equiv\Theta^a_{{\cV}_1{\cV}_2}\gamma^a=\hat{\Theta}_{{\cV}_1{\cV}_2}^{\dag}.
\label{Dirac_Form}
\end{gather}
Here each $\sigma$ is one of 5! vertex permutations ${\cV}_{({\cW})i}\longrightarrow\sigma(
{\cV}_{({\cW})i})$. It is assumed that $l^2_P$ is a real number.
It is easy to check that (compare with (\ref{Variables_Grav}))
\begin{gather}
\hat{\Theta}_{{\cV}_1{\cV}_2}
\equiv-\Omega_{{\cV}_1{\cV}_2}\hat{\Theta}_{{\cV}_2{\cV}_1}
\Omega_{{\cV}_1{\cV}_2}^{-1}.
\label{Dir_Bil_Form_Trans}
\end{gather}

The action (\ref{Latt_Action}) is invariant relative to the gauge transformations
\begin{gather}
\tilde{\Omega}_{{\cV}_1{\cV}_2}
=S_{{\cV}_1}\Omega_{{\cV}_1{\cV}_2}S^{-1}_{{\cV}_2}, \quad
\tilde{\hat{e}}_{{\cV}_1{\cV}_2}=S_{{\cV}_1}\,\hat{e}_{{\cV}_1{\cV}_2}\,S^{-1}_{{\cV}_1}, \quad
\tilde{\Psi}_{\cV}=S_{\cV}\Psi_{\cV}, \quad \tilde{\Psi^{\dag}}_{\cV}=\Psi_{\cV}^{\dag}S_{\cV}^{-1}, \quad  S_{\cV}\in\Spin(4).
\label{Gauge_Trans}
\end{gather}
Verification of this fact is facilitated by using the relation (compare with the relation for
$\hat{e}_{{\cV}_1{\cV}_2}$ in (\ref{Gauge_Trans}))
\begin{gather}
\tilde{\hat{\Theta}}_{{\cV}_1{\cV}_2}=S_{{\cV}_1}\hat{\Theta}_{{\cV}_1{\cV}_2}S^{-1}_{{\cV}_1},
\label{Teta_Gauge_Trans}
\end{gather}
which follows directly from (\ref{Gauge_Trans}).

When calculating the partition function (transition amplitude), it is more convenient to use an action that includes external sources:
\begin{gather}
\mA_{tot}=\mA+\Delta\mA_J,
\nonumber \\
\Delta{\mA}_J=\sum_{({\cV}_1{\cV}_2)}\tr
\hat{e}_{{\cV}_1{\cV}_2}J^{(e)}_{{\cV}_1{\cV}_2}+
\sum_{\cV}\left(J^{(\Psi)\dag}_{\cV}\Psi_{\cV}+\Psi_{\cV}^{\dag}J^{(\Psi)}_{\cV} \right).
\label{Total_Action}
\end{gather}
For the overall action $\mA_{tot}$ to be gauge invariant, external sources must be transformed
together with the transformation of variables (\ref{Gauge_Trans}) according to
\begin{gather}
\tilde{J}^{(e)}_{{\cV}_1{\cV}_2}=S_{{\cV}_1}J^{(e)}_{{\cV}_1{\cV}_2}S_{{\cV}_1}^{-1}, \quad
\tilde{J}^{(\Psi)\dag}_{\cV}=J^{(\Psi)\dag}_{\cV}S_{{\cV}}^{-1}, \quad
\tilde{J}^{(\Psi)}_{\cV}=S_{{\cV}}J^{(\Psi)}_{\cV}.
\label{Gauge_Trans_Ext}
\end{gather}

The partition function  is defined as follows:
\begin{gather}
\mU\{J\}=\int\prod_{\cV}(\d\Psi^{\dag}_{\cV}\d\Psi_{\cV})\prod_{({\cV}_1{\cV}_2)}(\d\Omega_{{\cV}_1{\cV}_2})
\left(\wedge\d e_{{\cV}_1{\cV}_2}\right)\exp(i\mA_{tot}).
\label{Partition_Function}
\end{gather}
Here $\d\Omega_{{\cV}_1{\cV}_2}$ is invariant measure on the group $\Spin(4)$,
\begin{gather}
\d e_{{\cV}_1{\cV}_2}=
\d e^1_{{\cV}_1{\cV}_2}\wedge\d e^2_{{\cV}_1{\cV}_2}\wedge\d e^3_{{\cV}_1{\cV}_2}
\wedge\d e^4_{{\cV}_1{\cV}_2}.
\nonumber
\end{gather}

We emphasize that in the integral (\ref{Partition_Function}) the action $\mA_{tot}$ is a real quantity the sign of which depends on the configuration of the variables. The imaginary unit in the exponent is introduced for the convergence of the integral.

The general meaning of studying a discrete model of gravity is to extract physical consequences from the assumption that space-time is discrete on an extremely small scale.

It seems that the lattice should not be "frozen". This means that the full lattice partition function must also include lattice summation. This raises difficult questions. For example, over what simplicial complexes should one sum up and with what weight? Should one sum over simplicial complexes without looking at their dimension and topology? We have no answer to these questions. However, as shown, in any case, in the long-wavelength limit, the memory of the lattice details is erased. But there are some physical consequences of the existence of granularity of space-time on ultra-small scales. For example, the doubling of fermion quanta in the sense of Wilson
\cite{vergeles2015wilson}, the existence of an instanton solution in the theory of gravity with topology $\R^4$
\cite{vergeles2017note}.

It is also interesting that any gauge theory on a lattice  generates a sum over surfaces as a result of high-temperature expansion. On the other hand, the sum over surfaces represents the propagation of a quantum string. This raises the question: are the assumption of the granularity of space-time on ultra-small scales and superstring theory dual descriptions of the same physics?

It should be noted that the first work on lattice gravity belongs to T. Regge \cite{regge1961general}.
Currently, there are many approaches to the discrete theory of gravity (see \cite{diakonov2011towards,
vladimirov2012phase,hamber2009quantum} and references there).

We also make the following remark. The wide fluctuations of the tetrad at short scale, which are discussed here, may be related to the spinfoam discreteness of geometrical quantities (See \cite{bianchi2013spinfoam,
engle2008lqg,perez2013spin}).  This connection is worth exploring, as the methods and approaches in one theory can enrich the other.

The organization of the paper is as follows.
Section II discusses the problem of convergence of the inner integral in (\ref{Partition_Function}).
In Section III a formal transition from lattice action to long-wavelength continual action is made. In addition, a transition is made from the Euclidean signature to the Minkowski signature.
In the next section, we define the global discrete symmetry for the lattice action, which is analogous to the PT transformation in the flat Minkowski space. The PT transformation defined on the lattice is interesting in that in the long-wavelength limit this transformation does not affect the emerging local coordinates (as is the case in the Minkowski space); it is realized exclusively on the dynamic variables of the theory.
In Section V the small scale quantum vacuum structure
is studied. It is shown that (i) the fluctuation values of tetrads $e_{{\cV}_1{\cV}_2}$ are commensurable with theirs most probable values; (ii) the macroscopic Universe can exist only in the presence of fermion fields.

\section{The convergence problem  of the integral (\ref{Partition_Function})}

\subsection{3D simplicial complex}

We first consider the convergence problem of the integral (\ref{Partition_Function}) in the pure theory of  gravity
$(\Psi=0)$ defined on a 3D simplicial complex. In this case we have (compare with (\ref{Variables_Grav}))
\begin{gather}
\Omega_{{\cV}_1{\cV}_2}=\Omega^{-1}_{{\cV}_2{\cV}_1}
=\exp\left(\frac{i}{2}\omega^{\alpha}_{{\cV}_1{\cV}_2}\sigma^{\alpha}\right)\in\SU(2),
\nonumber \\
\hat{e}_{{\cV}_1{\cV}_2}\equiv
-\Omega_{{\cV}_1{\cV}_2}\hat{e}_{{\cV}_2{\cV}_1}\Omega_{{\cV}_1{\cV}_2}^{-1}
\equiv e^{\alpha}_{{\cV}_1{\cV}_2}\sigma^{\alpha}.
\label{Variables}
\end{gather}
Everywhere $\sigma^{\alpha}, \  \alpha=1,2,3$ are Pauli matrices.
Now the Levi-Civita symbol
$\varepsilon_{{\cV}_{({\cW})1}{\cV}_{({\cW})2}{\cV}_{({\cW})3}{\cV}_{({\cW})4}}=\pm 1$ depending on
whether the order of vertices
$s^3_{\cW}=a_{{\cV}_{({\cW})1}}a_{{\cV}_{({\cW})2}}a_{{\cV}_{({\cW})3}}a_{{\cV}_{({\cW})4}}$ defines the
positive or negative orientation of 3-simplex $s^3_{\cW}$.
Introduce the notation for the lattice curvature:
\begin{gather}
{\mC}_{{\cV}_{({\cW})1}{\cV}_{({\cW})2}{\cV}_{({\cW})3}}
\equiv\Omega_{{\cV}_{({\cW})1}{\cV}_{({\cW})2}}
\Omega_{{\cV}_{({\cW})2}{\cV}_{({\cW})3}}
\Omega_{{\cV}_{({\cW})3}{\cV}_{({\cW})1}}\in\SU(2),
\nonumber \\
{\mC}_{{\cV}_{({\cW})1}{\cV}_{({\cW})2}{\cV}_{({\cW})3}}^{\alpha}
\equiv-\frac{i}{2}\tr\sigma^{\alpha}{\mC}_{{\cV}_{({\cW})1}{\cV}_{({\cW})2}{\cV}_{({\cW})3}}.
\label{CurvatureLat}
\end{gather}
The lattice pure gravity action with  external sources has the form
\begin{gather}
\mA_g=\frac{1}{4\cdot
6\cdot2\cdot l_P}\sum_{\cW}
\sum_{\sigma}\sum_{\alpha}
\varepsilon_{\sigma({\cV}_{({\cW})1})\sigma({\cV}_{({\cW})2})
\sigma({\cV}_{({\cW})3})\sigma({\cV}_{({\cW})4})}
{\mC}_{\sigma(
{\cV}_{({\cW})4})\sigma({\cV}_{({\cW})1})\sigma({\cV}_{({\cW})2})}^{\alpha}
e^{\alpha}_{\sigma({\cV}_{({\cW})4})\sigma({\cV}_{({\cW})3})}
\nonumber \\
+\sum_{\alpha,({\cV}_1{\cV}_2)}e^{\alpha}_{{\cV}_1{\cV}_2}J^{(e)\alpha}_{{\cV}_1{\cV}_2}.
\label{Action3D}
\end{gather}
This action is invariant relative to the gauge transformations
\begin{gather}
\tilde{\Omega}_{{\cV}_1{\cV}_2}
=S_{{\cV}_1}\Omega_{{\cV}_1{\cV}_2}S^{-1}_{{\cV}_2}, \quad
\tilde{\hat{e}}_{{\cV}_1{\cV}_2}=S_{{\cV}_1}\,\hat{e}_{{\cV}_1{\cV}_2}\,S^{-1}_{{\cV}_1},
\quad \tilde{J}^{(e)\alpha}_{{\cV}_1{\cV}_2}\sigma^{\alpha}=
\left(S_{{\cV}_1}\sigma^{\alpha}\,S^{-1}_{{\cV}_1}\right)J^{\alpha}_{{\cV}_1{\cV}_2},
\nonumber \\
S_{\cV}\in\SU(2).
\label{Invariance}
\end{gather}

Since the action (\ref{Action3D}) is a linear functional of $\{e\}$, then the internal  integral over the variables
$\{e^{\alpha}_{{\cV}_1{\cV}_2}\}$ in (\ref{Partition_Function})
gives the product of ${\mN}^{(1)}$ three-dimensional $\delta$-functions for each 1-simplex of the complex.
${\mN}^{(1)}$ denotes the number of 1-simplexes of a complex. Fix 1-simplex
$a_{{\cV}_4}a_{{\cV}_3}=a_{{\cV}_{({\cW})4}}a_{{\cV}_{({\cW})3}}\in s^3_{\cW}=a_{{\cV}_{({\cW})1}}a_{{\cV}_{({\cW})2}}
a_{{\cV}_{({\cW})3}}a_{{\cV}_{({\cW})4}}$, and the indicated arrangement of vertices determines a positive orientation
of 3-simplex $ s^3_{\cW}$. Using (\ref{Variables}) and (\ref{Action3D}), it is easy to find that integration over a variable
$e^{\alpha}_{{\cV}_4{\cV}_3}$ gives the following delta function as a factor of a partition function:
\begin{gather}
\prod_{\alpha=1,2,3}\delta\left\{\frac{1}{4\cdot 6\cdot l_P}\sum_{{\cW}:\, s^3_{\cW}\ni a_{{\cV}_4}a_{{\cV}_3}}\left[
{\mC}_{
{\cV}_{({\cW})4}{\cV}_{({\cW})1}{\cV}_{({\cW})2}}^{\alpha}+
\left(\Omega_{{\cV}_{({\cW})4}{\cV}_{({\cW})3}}{\mC}_{
{\cV}_{({\cW})3}{\cV}_{({\cW})1}{\cV}_{({\cW})2}}\Omega_{{\cV}_{({\cW})3}{\cV}_{({\cW})4}}
\right)^{\alpha}\right]+J^{\alpha}_{{\cV}_4{\cV}_3}\right\}.
\label{ContrPartFunc}
\end{gather}

We need to make some comments:

(i) Multiplicity of integration over variables $\{\Omega\}$ in (\ref{Partition_Function}) exactly corresponds to the number of one-dimensional delta functions which is equal to $3{\mN}^{(1)}$. Indeed, on each 1-simplex the independent variable
$\Omega\in\SU(2)$ is defined, and the group $\SU(2)$ is three-dimensional.

(ii) According to (\ref{ContrPartFunc}), a {\it fixed nonzero} external source violates gauge invariance.

The following conclusion follows from these comments: as a result of integration over variables $\{\Omega\}$  all $\delta$-functions (\ref{ContrPartFunc}) are integrated. For  $|J^{\alpha}_{{\cV}_1{\cV}_2}|\ll l_P^{-1}$, all delta functions make a nonzero contribution. On the contrary, if any $|J^{\alpha}_{{\cV}_1{\cV}_2}|\gg l_P^{-1}$, then the corresponding delta function remains zero, since its argument will always be nonzero. Therefore, the entire partition function turns out to be zero. Thus, the partition function is not an analytic function of external sources even for a finite simplicial complex. It is also obvious that for $J^{\alpha}_{{\cV}_1{\cV}_2}\longrightarrow 0$ the integral over  $\{\Omega\}$ is saturated only near two configurations: $\Omega_{{\cV}_1{\cV}_2}\sim1$
and $\Omega_{{\cV}_1{\cV}_2}\sim-1$. Only the first possibility is realized in the long-wavelength limit of the model.

Indeed, in the long-wavelength limit, the model (\ref{Action3D}) takes the form
\begin{gather}
\mA_g=\frac{1}{2\cdot l_P}\int\mR^{\alpha}_{\mu\nu}\d x^{\mu}\wedge\d x^{\nu}\wedge e^{\alpha}_{\lambda}\d x^{\lambda},
\nonumber \\
\mR_{\mu\nu}^{\alpha}=\partial_{\mu}\omega_{\nu}^{\alpha}-\partial_{\nu}\omega_{\mu}-\varepsilon_{\alpha\beta\gamma}
\omega_{\mu}^{\beta}\omega_{\nu}^{\gamma}.
\label{Action3DLong}
\end{gather}
After integration in the functional integral over the variables $\{e^{\alpha}_{\lambda}\}$ we get
$\mR_{\mu\nu}^{\alpha}=0$.

\subsection{4D simplicial complex}

Here we consider pure gravity ($\Psi=0$) with only $J^{(e)}_{{\cV}_1{\cV}_2}\neq0$.

Rewrite the action (\ref{Latt_Action_Grav}) in the form
\begin{gather}
\mA_{\mbox{tot}}=\frac{1}{2l_P^2}\sum_{a,b,\left({\cV}_1{\cV}_2\right),\left({\cV}_3{\cV}_4\right)}{\cal Q}^{ab}_{\left({\cV}_1{\cV}_2\right),\left({\cV}_3{\cV}_4\right)}\{\Omega\}
e^a_{{\cV}_1{\cV}_2}e^b_{{\cV}_3{\cV}_4}+
\sum_{a,({\cV}_1{\cV}_2)}e^a_{{\cV}_1{\cV}_2}J^{(e)a}_{{\cV}_1{\cV}_2}.
\label{quadratic}
\end{gather}
The kernel of a quadratic form ${\cal Q}\{\Omega\}$ is degenerate on the subspace of measure zero in the space of variables
$\{\Omega\}$. Indeed, this subspace is determined by the system the following equations
\begin{gather}
\det_{(ab)}{\mC}^{ab}_{{\cV}_1{\cV}_2{\cV}_3}=0,  \quad
{\mC}^{ab}_{{\cV}_1{\cV}_2{\cV}_3}
\equiv\tr\sigma^{ab}\Omega_{{\cV}_1{\cV}_2}
\Omega_{{\cV}_2{\cV}_3}
\Omega_{{\cV}_3{\cV}_1}
\nonumber
\end{gather}
on the space of degrees of freedom $\{\Omega\}$. Let ${\mN}^{(2)}$ be the number of 2-simplexes of the lattice.
Thus, if the dimension of the space of variables $\{\Omega\}$ is $6{\mN}^{(1)}$, then the dimension of the degeneration subspace of the  operator ${\cal Q}^{ab}_{\left({\cV}_1{\cV}_2\right),\left({\cV}_3{\cV}_4\right)}\{\Omega\}$ varies between the values
$\left(6{\mN}^{(1)}-1\right)$ and $\left(6{\mN}^{(1)}-{\mN}^{(2)}\right)$.

Hence, the integral over $\{e\}$ is a many dimensional Fresnel integral:
\begin{gather}
\int_e\cdot\exp\big(i\mA_{\mbox{tot}}\big)=\Const\cdot (l_P)^{4{\mN}^{(1)}}
\{\det{\cal Q}\}^{-1/2}\exp\left\{
-\frac{i\, l_P^2}{2}\sum_{a,b,\left({\cV}_1{\cV}_2\right),\left({\cV}_3{\cV}_4\right)}({\cal Q}^{-1})^{ab}_{\left({\cV}_1{\cV}_2\right),\left({\cV}_3{\cV}_4\right)}\{\Omega\}
J^{(e)a}_{{\cV}_1{\cV}_2}J^{(e)b}_{{\cV}_3{\cV}_4}\right\}.
\label{Fresnel}
\end{gather}
Here $\Const$ is a numerical one, it does not depend on either the parameters or the degrees of freedom of the system. Thus, the external $\Omega$ integral of (\ref{Fresnel}) must be convergent.

\section{Long wavelength limit and  Minkowski signature}

Now let us pass on to the limit of slowly varying fields, when the action
(\ref{Latt_Action}) reduces to the well known continuous gravity action. This transition have meaning
together with the transition to Minkowski signature. As a result the compact gauge group $\Spin(4)$ transforms into the non-compact group $\Spin(3,1)$.

In this section, all lattice quantities for the Euclidean signature are primed. For the quantities related to the Minkowski signature, the previous designations are retained. This is done to make the transition from the Euclidean signature to the Minkowski signature transparent.

Firstly let us perform the following deformations of integration contours  in integral (\ref{Partition_Function}):
\begin{gather}
{\omega'}_{{\cV}_1{\cV}_2}^{4\alpha}= i\omega^{0\alpha}_{{\cV}_1{\cV}_2}, \quad
{\omega'}_{{\cV}_1{\cV}_2}^{\alpha\beta}=-\omega_{{\cV}_1{\cV}_2}^{\alpha\beta},
\nonumber \\
{e'}^4_{{\cV}_1{\cV}_2}= e^0_{{\cV}_1{\cV}_2}, \quad {e'}^{\alpha}_{{\cV}_1{\cV}_2}= ie^{\alpha}_{{\cV}_1{\cV}_2}.
\label{Variables_Trans_Mink}
\end{gather}
The variables $\omega^{ab}_{{\cW}ij}$, $e^a_{{\cW}ij}$ are real quantities for Minkowski signature,
and the indices take on the values
\begin{gather}
a,\,b\ldots=0,1,2,3, \quad \alpha,\,\beta,\ldots=1,2,3.
\label{Mink_Ind}
\end{gather}
The Dirac matrices are transformed as follows:
\begin{gather}
{\gamma'}^4=\gamma^0, \quad {\gamma'}^{\alpha}= i\gamma^{\alpha}, \quad
{\gamma'}^5=\gamma^5=i\gamma^0\gamma^1\gamma^2\gamma^3,
\nonumber \\
\frac12(\gamma^a\gamma^b+\gamma^b\gamma^a)=\eta^{ab}=\diag(1,\,-1,\,-1,\,-1), \quad
\tr\gamma^5\gamma_a\gamma_b\gamma_c\gamma_d=4i\varepsilon_{abcd}, \quad \varepsilon_{0123}=1.
\label{Mink_Dirac_matr}
\end{gather}
Thus, for $\sigma^{ab}=(1/4)[\gamma^a,\,\gamma^b]$ we get
\begin{gather}
{\sigma'}^{4\alpha}= i\sigma^{0\alpha}, \quad {\sigma'}^{\alpha\beta}=-\sigma^{\alpha\beta}.
\label{Mink_Spin_Matr}
\end{gather}
Raising and lowering indices $a,b,\ldots$ is done using tensors $\eta^{ab}$ and $\eta_{ab}$, respectively.
 As a result of (\ref{Variables_Trans_Mink})-(\ref{Mink_Spin_Matr}) we have
\begin{gather}
{\omega'}_{{\cV}_1{\cV}_2}=\frac12\omega^{ab}_{{\cV}_1{\cV}_2}\,\sigma_{ab}
\equiv\omega_{{\cV}_1{\cV}_2},
\quad
{\hat{e}'}_{{\cV}_1{\cV}_2}=\gamma_ae^a_{{\cV}_1{\cV}_2}\equiv\hat{e}_{{\cV}_1{\cV}_2},
\label{Mink_5}
\end{gather}
and also
\begin{gather}
{\Omega'}_{{\cV}_1{\cV}_2}=\exp\left(\frac12{\omega'}_{{\cV}_1{\cV}_2}^{ab}{\sigma'}^{ab}\right)=
\exp\left(\frac12\omega_{{\cV}_1{\cV}_2}^{ab}\sigma_{ab}\right)\equiv\Omega_{{\cV}_1{\cV}_2}\in\Spin(3,1).
\label{Mink_6}
\end{gather}
We see that the holonomy elements $\Omega_{{\cV}_1{\cV}_2}$ become the elements of the non-compact group $\Spin(3,1)$.

Dirac variables are transformed according to
\begin{gather}
\Psi'_{\cV}=\Psi_{\cV}, \quad  {\Psi'}^{\dag}_{\cV}=\Psi_{\cV}^{\dag}\gamma^0=\overline{\Psi}_{\cV}.
\label{Mink_Dir}
\end{gather}
in passing to the Minkowski signature.

Now let us pass on to the formal limit of slowly varying fields, when the action
(\ref{Latt_Action}) reduces to the well known continuous gravity action.
In order to pass to the long-wavelength limit let's
consider a certain 4D sub-complex  ${\mK}'\in\mK$ with the trivial topology of four-dimensional disk.
Realize geometrically this sub-complex in $\R^4$.  Suppose that the
geometric realization is the triangulation of a compact part of $\R^4$, so that in $\R^4$ we have $(\mbox{int}\, s^4_{\cW})\cap (\mbox{int}\, s^4_{{\cW}'})=\emptyset$ for ${\cW}\neq{\cW}'$ and the sizes of these
simplices are commensurable. Thus each vertex of the sub-complex  acquires
the coordinates $x^{\mu}$  which are the coordinates of the vertex image in $\R^4$:
\begin{gather}
x^{\mu}_{\cV}\equiv x^{\mu}(a_{\cV}),
 \qquad \ \mu=1,\,2,\,3,\,4.
\label{intr110}
\end{gather}
We stress that the correspondence between the vertices
from the considered subset and the coordinates (\ref{intr110}) is one-to-one.

Consider a certain simplex  $s^4_{\cW}$. We denote all five vertices of this 4-simplex as ${\cV}_i$, $i=1,2,3,4$ and ${\cV}_m\neq {\cV}_i$.
Note that each of these vertices also belongs to other adjacent 4-simplexes.
From the aforesaid it is evident that the four infinitesimal vectors
\begin{gather}
\d x^{\mu}_{{\cV}_m{\cV}_i}\equiv x^{\mu}_{{\cV}_i}-x^{\mu}_{{\cV}_m}=
-\d x^{\mu}_{{\cV}_i{\cV}_m}\in\R^4 ,  \quad
i=1,\,2,\,3,\,4
\label{intr120}
\end{gather}
are linearly independent. The  vector system (\ref{intr120}) refers to the simplex $ s^4_{\cW}$ and its vertex
${\cV}_m$, while each of the vectors of
the system (\ref{intr120}) belongs to several adjacent 4-simplexes.
The differentials of coordinates
(\ref{intr120}) correspond to one-dimensional simplices $a_{{\cV}_m}a_{{\cV}_i}$, so that,
if the vertex $a_{{\cV}_m}$ has coordinates $x^{\mu}_{{\cV}_m}$, then the vertex
$a_{{\cV}_i}$ has the coordinates $x^{\mu}_{{\cV}_m}+\d x^{\mu}_{\left({\cV}_m{\cV}_i\right)}$.

In the continuous limit, the holonomy group elements  $\Omega_{{\cV}_1{\cV}_2}$ are
close to the identity element, so that the quantities $\omega^{ab}_{{\cV}_1{\cV}_2}$
tend to zero being of the order of $O(\d x^{\mu})$ and they slowly change  when moving along the lattice.
Thus one can consider the following system of equation for
$\omega_{\mu}\left(\cW,{\cV}_m,{\cV}_i\right)$:
\begin{gather}
\omega_{\mu}\left(\cW,{\cV}_m,{\cV}_i\right)\,\d x^{\mu}_{{\cV}_m{\cV}_i}=\omega_{
{\cV}_{m}{\cV}_{i}},  \quad
i=1,\,2,\,3,\,4\,.
\label{intr130}
\end{gather}
In this system of linear equation, the indices ${\cW}$ and $m$ are
fixed, the summation is carried out over the index $\mu$. Since the vectors
(\ref{intr120}) are linearly independent, the quantities $\omega_{\mu}\left(\cW,{\cV}_m,{\cV}_i\right)$
are defined uniquely.

The problem is  the definition of the connection 1-form field
$\omega_{\mu}\big((x_{{\cV}_m}+x_{{\cV}_i})/2\big)$
satisfying systems of equations (\ref{intr130}) for all 1-simplexes $a_{{\cV}_m}a_{{\cV}_i}$
independently of 4-simplexes and the vertices marked in them. The inverse problem is solved obviously:
if a 1-form  $\omega_{\mu}\big((x_{{\cV}_m}+x_{{\cV}_i})/2\big)$ is given, then the set of values $\omega_{
{\cV}_{m}{\cV}_{i}}$ is determined according to (\ref{intr130}).

Note that due to Eqs. (\ref{Variables_Grav}), (\ref{intr120}) and (\ref{intr130}) we have
\begin{gather}
\omega_{\mu}\left(\cW,{\cV}_m,{\cV}_i\right)\,\d x^{\mu}_{{\cV}_m{\cV}_i}=
\omega_{\mu}\left(\cW,{\cV}_i,{\cV}_m\right)\,\d x^{\mu}_{{\cV}_m{\cV}_i}.
\label{intr132}
\end{gather}
Suppose that a one-dimensional simplex $a_{{\cV}_m}a_{{\cV}_i}$
belongs to four-dimensional simplices with indices
${\cW}_1,\,{\cW}_2,\,\ldots\,,\,{\cW}_r$. Introduce the quantity
\begin{gather}
\omega_{\mu}\left(\frac{1}{2}\,(x_{{\cV}_m}+
x_{{\cV}_i})\,\right)
\equiv\frac{1}{2r}\sum_{s=1}^r
\bigg\{\omega_{\mu}\left({\cW}_s,{\cV}_m,{\cV}_i\right)+\omega_{\mu}\left({\cW}_s,{\cV}_i,{\cV}_m\right)\bigg\}\,,
\label{intr140}
\end{gather}
which is assumed to be related to the midpoint of the segment
$[x^{\mu}_{{\cV}_m},\,x^{\mu}_{{\cV}_i}\,]$.  According to the
definition, we have the following chain of identities:
\begin{gather}
\omega_{{\cV}_{({\cW}_1)m}{\cV}_{({\cW}_1)i}}
\equiv\ldots\equiv\omega_{{\cV}_{({\cW}_r)m}{\cV}_{({\cW}_1)r}}.
\label{intr150}
\end{gather}
It follows from (\ref{intr120})--(\ref{intr150}) that
\begin{gather}
\omega_{\mu}\left(\frac{1}{2}\,(x_{{\cV}_m}+
x_{{\cV}_i})\,\right)\,\d
x^{\mu}_{{\cV}_m{\cV}_i}=\omega_{{\cV}_m{\cV}_i}  \,.
\label{intr160}
\end{gather}
The value of the field element $\omega_{\mu}$ in (\ref{intr160}) is uniquely defined by the  midpoint of corresponding
one-dimensional simplex. At any points, the connection 1-form is determined by continuity.

Hence, we assume that the fields $\omega_{\mu}$ smoothly depend on
the points belonging to the geometric realization of each
four-dimensional simplex. In this case, the following formula is
valid up to $O\big((\d x)^2\big)$ inclusive
\begin{gather}
\Omega_{{\cV}_m{\cV}_i}\Omega_{{\cV}_i{\cV}_j}\Omega_{{\cV}_j{\cV}_m}=
\exp\left[\frac{1}{2}\,\mR_{\mu\nu}(x_{{\cV}_m})\d x^{\mu}_{\left({\cV}_m{\cV}_i\right)} \d
x^{\nu}_{\left({\cV}_m{\cV}_j\right)}\right],
\label{Lat_curv}
\end{gather}
where
\begin{gather}
\mR_{\mu\nu}=\partial_{\mu}\omega_{\nu}-\partial_{\nu}\omega_{\mu}+
[\omega_{\mu},\,\omega_{\nu}\,]\,.
\label{Long_W_Curv}
\end{gather}

In exact analogy with (\ref{intr130}), let us write out the following relations
for a tetrad field without explanations:
\begin{gather}
\hat{e}_{\mu}\left(\cW,{\cV}_m,{\cV}_i\right)\d x^{\mu}_{{\cV}_m{\cV}_i}=
\hat{e}_{{\cV}_m{\cV}_i}, \quad i=1,2,3,4.
\label{intr190}
\end{gather}
So 1-form
\begin{gather}
e^a(x)=e^a_{\mu}(x)\d x^{\mu}
\label{intr192}
\end{gather}
arises.

With the help of Equations given in this Section, we rewrite the lattice action (\ref{Latt_Action}) in the long-wavelength limit:
\begin{gather}
i{\mA'}_{g\,Lat}\longrightarrow\mA_g,  \quad \mA_g=-\frac{1}{4\,l^2_P}\varepsilon_{abcd}\int\mR^{ab}\wedge e^c\wedge e^d,  \quad
\mR^{ab}=\mR^{ab}_{\mu\nu}\d x^{\mu}\wedge\d x^{\nu},
\label{Long_Wav_Grav_Act}
\end{gather}
\begin{gather}
i{\mA'}_{\Psi\,Lat}\longrightarrow\mA_{\Psi}, \quad \mA_{\Psi}=\frac16\varepsilon_{abcd}\int\Theta^a\wedge e^b\wedge e^c\wedge e^d,
\nonumber \\
\Theta^a=\frac{i}{2}\left[\overline{\Psi}\gamma^a{\cal D}_{\mu}\,\Psi-
\left(\overline{{\cal D}_{\mu}\,\Psi}\right)\gamma^a\Psi\right]\d x^{\mu}, \quad
{\cal D}_{\mu}=\left(\partial_{\mu}+\omega_{\mu}\right).
\label{Long_Wav_Dir_Act}
\end{gather}
In formulas (\ref{Long_Wav_Grav_Act})-(\ref{Long_wave_Minkowski}) the subscript $"Lat"$ means that the corresponding values are defined on the lattice.
Note that as a result of the indicated deformation of the integration contours, the quantities (\ref{Long_Wav_Grav_Act}) and (\ref{Long_Wav_Dir_Act})  became purely real.
The transformation of these quantities into a purely imaginary values is done by converting one of the coordinates into a purely imaginary coordinate:
$x^4=ix^0$, and the coordinate $x^0$ is considered real. In what follows, in the case of the Minkowski signature, the local coordinates are denoted as $x^{\mu},\,\mu=0,\,1,\,2,\,3$. Thus the transformation $\mA=\mA_g+\mA_{\Psi}\longrightarrow i\mA$ is realized. Since the forms $\omega^{ab}_{\mu}$ and $e^a_{\mu}$  are considered to be real fields and the differentials $\d x^4=i\d x^0$ are purely imaginary, then according to the Eqs. (\ref{intr130}) and (\ref{intr190}) the lattice variables $\omega_{{\cV}_m{\cV}_i}^{ab}$ and $e^a_{{\cV}_m{\cV}_i}$ become complex. In other words, the corresponding deformation of the integration contours over lattice variables into the complex plane gives the necessary effect.
We have also $\Theta^a_4=\Theta^a_0$. This conclusion follows from the fact that the value $\partial_{\mu}\Psi$ in (\ref{Long_Wav_Dir_Act}) is obtained from the difference $(\Psi_{{\cV}_i}-\Psi_{{\cV}_m})$.
So the subscript $4$ only indicates the direction of the field $\Psi$ shift.

As a result of the substitution $x^4=ix^0$, the functional $\mA$ is replaced by $(i\mA)$, and the functional $\mA$ is real.

The above means that in the long-wavelength limit in the case of the Minkowski signature we have
\begin{gather}
\exp\left(i\mA'_{Lat}\right)\longrightarrow\exp(i\mA).
\label{Long_wave_Minkowski}
\end{gather}

The expressions (\ref{Long_Wav_Grav_Act}) and (\ref{Long_Wav_Dir_Act}) specify the Hilbert-Einstein-Dirac action in the Palatini form. This action is invariant relative to the gauge transformation
\begin{gather}
\tilde{\hat{e}}_{\mu}=S\hat{e}_{\mu}S^{-1}, \quad \tilde{\omega}_{\mu}=S\omega_{\mu}S^{-1}+S\partial_{\mu}S^{-1},
\quad
\tilde{\Psi}=S\Psi, \quad \tilde{\Psi^{\dag}}=\Psi^{\dag}S^{-1}, \quad S\in\Spin(3,1).
\label{Gage_Transf_Long_W_Min}
\end{gather}

We also write out the long wavelength limit of  action (\ref{Latt_Action}) in the case of the Euclidean signature:
\begin{gather}
\mA_{Long}'=\frac{1}{4\,l^2_P}\varepsilon_{abcd}\int{\mR'}^{ab}\wedge {e'}^c\wedge {e'}^d-\frac16\varepsilon_{abcd}\int{\Theta'}^a\wedge {e'}^b\wedge {e'}^c\wedge {e'}^d,
\label{Long_W_Eucl_Ac}
\end{gather}
where ${\mR'}^{ab}_{\mu\nu}$ is given by (\ref{Long_W_Curv}) for ${\omega'}_{\mu}=(1/2){\omega'}_{\mu}^{ab}{\sigma'}^{ab}$ and
\begin{gather}
{\Theta'}^a_{\mu\,Long}=\frac{i}{2}\left[{\Psi'}^{\dag}{\gamma'}^a{\cal D}'_{\mu}\,\Psi'-
\left({{\cal D}'_{\mu}\,\Psi}'\right)^{\dag}{\gamma'}^a\Psi'\right], \quad
{\cal D}'_{\mu}=\left(\partial_{\mu}+\omega'_{\mu}\right).
\nonumber
\end{gather}
it is obvious that
\begin{gather}
\exp\left(i\mA'_{Lat}\right)\longrightarrow\exp(i\mA_{Long}').
\label{Long_wave_Eu_Trans}
\end{gather}
In  (\ref{Long_W_Eucl_Ac}) the local coordinates $x^{\mu}$ (\ref{intr110}) are used.
The action (\ref{Long_W_Eucl_Ac}) is invariant relative to the gauge transformation
(\ref{Gage_Transf_Long_W_Min}) for $S\in\Spin(4)$.

The relation (\ref{Long_wave_Minkowski}) shows the following: Suppose that in the theory of quantum gravity on a lattice with a Euclidean signature, an oscillating exponent is used in the integral for the partition functin  (\ref{Partition_Function}). Then, due to only deformations of the integration contours, it is possible to go over to the Minkowski signature and at the same time have an oscillating exponent in the integral that determines the transition amplitude.

\section{Lattice variant of PT-symmetry}

Consider the global discrete transformation of the degrees of freedom in the lattice theory in the framework of the Euclidean signature. Further, the primed variables are not used, since this does not lead to confusion. The transformed variables are denoted by the superscript $"PT"$, their explicit expressions through the original variables is as follows:
\begin{gather}
\left(\Psi^{PT}\right)_{\cV}=U_{PT}\left(\Psi^{\dag}_{\cV}\right)^t, \quad
\left(\Psi^{{\dag}PT}\right)_{\cV}=-\left(\Psi_{\cV}\right)^tU^{-1}_{PT},
\nonumber \\
\left(e^{PT}\right)^a_{{\cV}_1{\cV}_2}=-e^{a}_{{\cV}_1{\cV}_2}, \quad
\left(\omega^{PT}\right)^{ab}_{{\cV}_1{\cV}_2}=\omega^{ab}_{{\cV}_1{\cV}_2},
\nonumber \\
 U_{PT}=\gamma^1\gamma^3.
\label{PT_transform}
\end{gather}
Here the superscript $"t"$ denotes the matrix transposition of the Dirac matrices and spinors.
We use a representation of the Dirac matrices in which
\begin{gather}
\left(\gamma^{1,3}\right)^t=-\gamma^{1,3},  \quad  \left(\gamma^{2,4}\right)^t=\gamma^{2,4}.
\nonumber
\end{gather}
Therefore, we have:
\begin{gather}
U^{-1}_{PT}\gamma^aU_{PT}=(\gamma^a)^t, \quad
U^{-1}_{PT}\sigma^{ab}U_{PT}=-(\sigma^{ab})^t.
\label{PT_trans_Dir_Alg}
\end{gather}
It follows from (\ref{PT_transform}) and (\ref{PT_trans_Dir_Alg}) that
\begin{gather}
 U^{-1}_{PT}\Omega_{{\cV}_1{\cV}_2}U_{PT}=\left(\Omega_{{\cV}_2{\cV}_1}\right)^t.
\label{PT_trans_Conn}
\end{gather}
Now, using (\ref{PT_transform}) and (\ref{PT_trans_Conn}) we find (compare with the tetrad transformation
 $e^a_{{\cV}_1{\cV}_2}$ in (\ref{PT_transform})):
\begin{gather}
\left(\Theta^{PT}\right)^a_{{\cV}_1{\cV}_2}=-\Theta^a_{{\cV}_1{\cV}_2}.
\label{PT_trans_Dir_Ac}
\end{gather}
In the process of checking the last relation, the fact that the spinor Fermi variables are anticommutative was taken into account.

According to (\ref{PT_transform}) and (\ref{PT_trans_Dir_Ac}) the action (\ref{Latt_Action}) is invariant
relative to the global discrete transformation of variables (\ref{PT_transform}).

In the case of the long wavelength limit and Euclidean signature the action (\ref{Long_W_Eucl_Ac})
is invariant relative to the global discrete transformation
\begin{gather}
\left(\Psi^{PT}\right)(x)=U_{PT}\left(\Psi^{\dag}(x)\right)^t, \quad
\left(\Psi^{{\dag}PT}\right)(x)=-\left(\Psi(x)\right)^tU^{-1}_{PT},
\nonumber \\
\left(e^{PT}\right)^a_{\mu}(x)=-e^{a}_{\mu}(x), \quad
\left(\omega^{PT}\right)^{ab}_{\mu}(x)=\omega^{ab}_{\mu}(x).
\label{PT_trans_Cont}
\end{gather}
Indeed, since the relations (\ref{PT_trans_Dir_Alg}) remain valid, then
\begin{gather}
\left(\Theta^{PT}\right)^a_{\mu}(x)=-\Theta^a_{\mu}(x),
\label{PT_trans_Cont_2}
\end{gather}
and therefore the action (\ref{Long_W_Eucl_Ac}) is invariant.

In the case of the Minkowski signature, the action ((\ref{Long_Wav_Grav_Act})+(\ref{Long_Wav_Dir_Act})) is invariant under the following global discrete transformation of variables:
\begin{gather}
\left(\Psi^{PT}\right)(x)=U_{PT}\left(\overline{\Psi}(x)\right)^t, \quad
\left(\overline{\Psi}^{PT}\right)(x)=-\left(\Psi(x)\right)^tU^{-1}_{PT},
\nonumber \\
\left(e^{PT}\right)^a_{\mu}(x)=-e^{a}_{\mu}(x), \quad
\left(\omega^{PT}\right)^{ab}_{\mu}(x)=\omega^{ab}_{\mu}(x),
\nonumber \\
U^{PT}=i\gamma^3\gamma^1.
\label{PT_trans_Cont_Minkovski}
\end{gather}
This fact is verified using equalities (\ref{PT_trans_Dir_Alg}), which are preserved in the case of the Minkowski signature.

Obviously discrete transformations (\ref{PT_transform}), (\ref{PT_trans_Cont}) or (\ref{PT_trans_Cont_Minkovski}) are not contained in the gauge transformation group (\ref {Gauge_Trans})
or (\ref{Gage_Transf_Long_W_Min}). Indeed, the described $PT$-transformations mutually replace
the fields $\Psi$ and $\Psi^{\dag}$, that is, they mutually replace particles and antiparticles.
On the contrary, for the
gauge transformations, the field $\Psi$ is transformed only through the field $\Psi$, and the field
$\Psi^{\dag}$ --- through the field $\Psi^{\dag}$.

It is easy to check that repeated application of the $PT$ -transformation operation gives the identical transformation. Discrete transformations (\ref{PT_transform}), (\ref{PT_trans_Cont}) or (\ref{PT_trans_Cont_Minkovski}) commute with gauge transformations.

We emphasize that the transformation of variables (\ref{PT_trans_Cont_Minkovski}) is qualitatively different from combined transformation of inversion and time reversal in flat Minkowski space.
In the Minkowski space, we have $e^a_{\mu}(x)=\delta^a_{\mu} $, $\omega^{ab}_{\mu}=0$ and the coordinates $x^{\mu}$ are global cartesian coordinates. Combined $PT$-symmetry in Minkowski Space
(for the Fermi part of the action) is determined by transformations of fermionic fields according to
(\ref{PT_trans_Cont_Minkovski}) and reflection of coordinates $(x^{PT})^{\mu}=-x^{\mu}$.
However, since in a curved space-time, generally speaking, only local coordinates are possible, which
can be chosen absolutely arbitrarily, it seems to us that
it is impossible to give a global meaning to the operations of reflection of local coordinates.
Moreover, in the definition of the lattice version of the theory, the local coordinates
absent; they appear only when going to the long-wavelength limit as vertex indices ("markers").
Therefore, in our opinion, discrete transformations (\ref{PT_transform}), (\ref{PT_trans_Cont}) or (\ref{PT_trans_Cont_Minkovski}) should be used in the theory of gravity, for which
retained the designation $"PT"$. Note that in the lattice version of the theory, the reflection $e\longrightarrow-e$ is consistent with the (\ref{Variables_Grav}) rule comparing $e^a_{{\cV}_1{\cV}_2}$ and $e^a_{{\cV}_2{\cV}_1}$. In the long-wavelength limit, the PT reflection of the tetrad commutes with the operation of changing the local coordinates, since when the local coordinates change, the tetrad is transformed according to the linear law:
$e^a_{\mu'}=(\partial x^{\mu}/\partial x^{\mu'})e^a_{\mu}$.

Further in this Section, we consider the long-wavelength limit in the case of the Minkowski signature.

From the above, the conclusion follows: if for $J\longrightarrow0$ there is a certain state $|\Lambda \rangle$,
then there is also a state $|\Lambda; PT \rangle$, obtained from the first by means of the $ PT $-transformation.
Let
\begin{gather}
\langle\Lambda|e^a_{\mu}(x)|\Lambda\rangle={\me}^a_{\mu}(x), \quad
\langle\Lambda|\Theta^a_{\mu}(x)|\Lambda\rangle=\theta^a_{\mu}(x).
\label{Mean}
\end{gather}
Then, according to (\ref{PT_trans_Cont_2}) and (\ref{PT_trans_Cont_Minkovski})
\begin{gather}
\langle\Lambda|\left(e^{PT}\right)^a_{\mu}(x)|\Lambda\rangle=-\me^a_{\mu}(x), \quad
\langle\Lambda|\left(\Theta^{PT}\right)^a_{\mu}(x)|\Lambda\rangle=-\theta^a_{\mu}(x).
\label{Mean_PT_var}
\end{gather}
On the other hand
\begin{gather}
\langle\Lambda;PT|\left(e^{PT}\right)^a_{\mu}(x)|\Lambda;PT\rangle=\me^a_{\mu}(x), \quad
\langle\Lambda;PT|\left(\Theta^{PT}\right)^a_{\mu}(x)|\Lambda;PT\rangle=\theta^a_{\mu}(x),
\label{Mean_PT_State_var}
\end{gather}
\begin{gather}
\langle\Lambda;PT|e^a_{\mu}(x)|\Lambda;PT\rangle=-\me^a_{\mu}(x), \quad
\langle\Lambda;PT|\Theta^a_{\mu}(x)|\Lambda;PT\rangle=-\theta^a_{\mu}(x).
\label{Mean_PT_State}
\end{gather}

Obviously, the sign of $\langle\Theta^a_{\mu}\rangle$ is determined by the filling rule of the Dirac
sea. To clarify the situation, consider the Dirac part of the energy-momentum tensor which has the form
\begin{gather}
t^{\mu\nu}_{\Psi}=\frac12\left\{\tilde{e}^{\mu}_a\Theta^{a\nu}+
\tilde{e}^{\nu}_a\Theta^{a\mu}\right\}-g^{\mu\nu}{\cal L}_{\Psi},  \quad {\cal L}_{\Psi}=\tilde{e}^{\mu}_a\Theta^a_{\mu}.
\label{Dirac_Energy_Mom_Ten}
\end{gather}
Here $\tilde{e}^{\mu}_ae_{\mu}^b=\delta^b_a$ and $g^{\mu\nu}=\eta^{ab}\tilde{e}^{\mu}_a\tilde{e}^{\nu}_b$ is the inverse metric tensor, ${\cal L}_{\Psi}$ is the the Lagrangian density.

Consider the case $g_{0i}=0$, $e^0_i=0$, $e^{\alpha}_0=0$. Then the conserved quantities (if we add to them
pseudotensor of the gravitational field) can be written as
\begin{gather}
P_{\mu}=\frac16\int_{\Sigma}\sqrt{-g}\varepsilon_{\nu\lambda\rho\sigma}t^{\ \ \nu}_{\Psi\,\mu}
\d x^{\lambda}\wedge\d x^{\rho}\wedge\d x^{\sigma}.
\label{Dirac_Grav_Ham}
\end{gather}
If the hypersurface $\Sigma$ is $x^0=\const$, then the zero component in
(\ref{Dirac_Grav_Ham}) turn into Dirac Hamiltonian
\begin{gather}
{\cal H}_{\Psi}=-\frac12\varepsilon_{\alpha\beta\gamma}\int_{x^0=\const}e^0_0\cdot\Theta^{\alpha}\wedge
e^{\beta}\wedge e^{\gamma}.
\label{Dirac_Grav_Simpl_Ham}
\end{gather}
Recall that $\Theta^{\alpha}\equiv\Theta^{\alpha}_i\d x^i$, $e^{\alpha}\equiv e^{\alpha}_i\d x^i$.
From (\ref{Dirac_Energy_Mom_Ten})-(\ref{Dirac_Grav_Simpl_Ham}) it is obvious that the reflection
$e\longrightarrow-e$ entails the need to redefinition the Dirac ground state by mutual replacement
particles and antiparticles. The PT transformation (\ref{PT_trans_Cont_Minkovski}) leaves
the Hamiltonian (\ref{Dirac_Grav_Simpl_Ham})  invariant.

In the case $e^{\mu}_a\longrightarrow\delta^{\mu}_a$
\begin{gather}
{\cal H}_{\Psi}=-\int\d^{(3)}x\sum_{\alpha=1}^3\Theta^{\alpha}_{\alpha}
=\int\d^{(3)}x\,\Psi^{\dag}\left(-i\gamma^0\gamma^{\alpha}\partial_{\alpha}\right)\Psi.
\label{Minkovski_Hamiltonian}
\end{gather}
The spectrum of the operator $\left(-i\gamma^0\gamma^{\alpha}\partial_{\alpha}\right)$ has both positive
$|{\bf k}|$ and negative $(-|{\bf k}|)$ eigenvalues, and there is
one-to-one correspondence between positive and negative frequency wave functions.
In the ground state of the system, all levels with negative energy of this operator are filled (Dirac sea), so that all excitations of the Hamiltonian (\ref{Minkovski_Hamiltonian}) turn out to be positive-frequency.
Let us denote the corresponding ground state as $|0\rangle$. Obviously
\begin{gather}
\langle 0|\Theta^{\alpha}_{\alpha}|0\rangle=2\int\frac{\d^{(3)}k}{(2\pi)^3}|{\bf k}|\longleftrightarrow
\langle 0|{\cal H}_{\Psi}|0\rangle=-2V\int\frac{\d^{(3)}k}{(2\pi)^3}|{\bf k}|.
\label{Mean_Vac}
\end{gather}
The $PT$-transformation swaps the positions of the positive and negative frequency wave functions.
This can be seen from the chain of equations, in which each next equation follows from the previous one:
\begin{gather}
-i\gamma^0\gamma^{\alpha}\partial_{\alpha}\psi_N=\epsilon_N\psi_N\longrightarrow
i\partial_{\alpha}\overline{\psi}_N\gamma^{\alpha}\gamma^0=\epsilon_N\overline{\psi}_N
\longrightarrow-i\gamma^0\gamma^{\alpha}\partial_{\alpha}\psi_N^{PT}=(-\epsilon_N)\psi_N^{PT}.
\label{Dirac_Mode_Trans}
\end{gather}
Therefore, the state $|0;PT\rangle$ is constructed by filling all levels with positive energy of the
operator $\left(-i\gamma^0\gamma^{\alpha}\partial_{\alpha}\right)$. We also have
$(\Theta^{PT})^a_{\mu}=-\Theta^a_{\mu}$, $(e^{PT})^{\mu}_a=-\delta^{\mu}_a$, so that
\begin{gather}
{\cal H}^{PT}_{\Psi}=\int\d^{(3)}x\, (t^{PT}_{\Psi})^{44}
=\int\d^{(3)}x\,(\Psi^{PT})^{\dag}\left(-i\gamma^0\gamma^{\alpha}\partial_{\alpha}\right)\Psi^{PT}.
\label{Hamiltonian_PT}
\end{gather}
According to (\ref{Dirac_Mode_Trans}) and (\ref{Hamiltonian_PT}) we again have
\begin{gather}
\langle 0;PT|{\cal H}^{PT}|0;PT\rangle=-2V\int\frac{\d^{(3)}k}{(2\pi)^3}|{\bf k}|,
\label{Mean_Vac_Complete}
\end{gather}

Let's denote by $C^{\dag}$ a {\it local} combined creation operator which is not $PT$-invariant: $C\neq C^{PT}$.
The experimenter can prepare the states $|C\rangle=C^{\dag}|0\rangle$ and $|C^{PT}\rangle=(C^{PT})^{\dag}|0\rangle$,
but not the state $|C^{PT};\,PT\rangle=(C^{PT})^{\dag}|0;\,PT\rangle$ since the vacuum state of all Universe can not be changed by the efforts of experimenter. Therefore the global reflection (\ref{PT_trans_Cont_Minkovski}) is absent in the experiment, and the evolutions of the states  $|C\rangle$ and $|C^{PT}\rangle$ are not connected by the $PT$-transformation. But the evolutions of the states $|C\rangle$ and $|C^{PT};\,PT\rangle$ would be identical after replacing all  degrees of freedom by the corresponding PT-transformed degrees of freedom.

Thus we come to the following conclusion: Despite the fact that the full actions (\ref{Latt_Action}) or
(\ref{Long_Wav_Grav_Act})-(\ref{Long_Wav_Dir_Act}) are  PT-invariant,  the experimenter will record a violation of PT-parity. So, PT-invariance is violated in the theory of gravity coupled with Dirac or Weyl fields.
This situation takes place in Nature.

The described PT  symmetry breaking can be considered as a spontaneous symmetry breaking.

The problem of discrete PT-symmetry in gravity is studied intensively.  See e.g the works
\cite{volovik2003universe,nissinen2018dimensional,volovik1990superfluid,wetterich2012universality,
sexty2013emergent,salomaa1987quantized,salomaa1988cosmiclike,nissinen2017effective}.

As shown, the considered theory contains at least two ground states $|0 \rangle$ and $| 0; PT \rangle $, which mutually transform under the action of $PT$-symmetry. Suppose the
state $ | 0 \rangle $ (or $ | 0; PT \rangle $) is realised. Obviously, in any of these possibilities, the symmetry between particles and antiparticles will be broken. Thus, the question arises: can the inclusion of gravity lead to clarify the mechanism of baryon asymmetry?

Concerning the problem of baryon asymmetry here it should be noted that in the framework of the Standard Model, the effect of baryon asymmetry takes place, but it 9 orders of magnitude less than the observed \cite{sakharov1991violation,bernreuther2002cp,chupp2019electric}. It is also worth mentioning here the significant efforts made for a long time to experimental search for electric dipole moments (EDM) for some hadrons and atoms. General scientific interest in experiments on searching for EDM lies, depending on the experimental results, in the possible clarification of the mechanism of the onset of baryon asymmetry.

The problem of determining the $CPT$ -invariant ground state in the continuum theory of gravity, including Dirac fields, is studied in \cite{boyle2018c}. The same paper discusses the consequences of the operation
reflections of the tetrad $e\longrightarrow-e$.

\section{Vacuum structure}

The theory of gravity is a non-renormalizable theory. This means, in particular, that quantum fluctuations of the proper gravitational degrees of freedom decay very quickly in the long-wavelength limit.
In the case of pure gravity, the metric tensor $g_{\mu\nu}$ is sufficient to describe it in the long-wavelength limit. Below we present some of the results obtained in the work \cite{t1974one}. Since the metric tensor fluctuates weakly, it can be represented as
\begin{gather}
\overline{g}_{\mu\nu}=g_{\mu\nu}+l_Ph_{\mu\nu}.
\label{long250}
\end{gather}
In what follows $g_{\mu\nu}$  means classical field satisfying classical Einstein equation, and $h_{\mu\nu}$
is quantum field. The Hilbert-Einstein gravitational action depends on the full metric tensor $\overline{g}_{\mu\nu}$ in the usual way.  In the considered case we have the following form for the correlator:
\begin{gather}
\langle\, h(0)h(x)\,\rangle=\O(1/x^2).
\label{long300}
\end{gather}
Therefore according to Eqs. (\ref{long250}) and (\ref{long300}) the quantum corrections to the classical field $g_{\mu\nu}$ at the scale $\sim x\gg l_P$ are of the order of
\begin{gather}
\delta g_{\mu\nu}\sim(l_P/x), \quad
(\delta  g_{\mu\nu})/g_{\mu\nu}=\O(l_P/x)\longrightarrow0.
\label{long310}
\end{gather}
It follows from here that the classical theory of gravity is adequate model for long-wavelength scales since the metric quantum fluctuations can be neglected. It is known that the electromagnetic, Yang-Mills, Dirac fields
long-wavelength quantum fluctuations are more essential (most crucial).

On the contrary, quantum fluctuations of the gravitational field  reconstruct radically the theory at small scales (the lattice in our case). Further consideration in this section is carried out for the lattice version of the theory.

 Write out the average value of any field in the presence of external sources:
\begin{gather}
\langle\Phi\rangle=\frac{\delta}{\delta J}\ln\mU\{J\}.
\label{MeanField}
\end{gather}

Introduce the following notations:
\begin{gather}
\langle\Phi\rangle_g=\frac{1}{N}\int_{\Omega,\,e}\Phi\cdot\exp\big(\mA\big)\bigg|_{J\longrightarrow0}, \quad
N=\int_{\Omega,\,e}\exp\big(\mA\big)\bigg|_{J\longrightarrow0}.
\label{small21}
\end{gather}
Here $\Phi$ denotes any functional of the dynamic variables.
The designation $\langle\Phi\rangle_{g,\Psi=0}$ is the mean (\ref{small21}) in the case when the fermion field is absent.

It is very interesting to consider the  mean $\langle e^a\rangle_{g,\Psi=0}$ in the theory of pure gravity, when matter (fermions in our case) is absent. Evidently,
\begin{gather}
\left\langle e^a_{\cV_1\cV_2}\right\rangle_{g,\Psi=0}=0,
\label{small30}
\end{gather}
since
\begin{gather}
\mA_g\{e\}=\mA_g\{-e\}.
\label{small32}
\end{gather}
It follows from here that for the
mean-square deviation we have
\begin{gather}
\left\langle\left(\delta e^a_{\cV_1\cV_2}\right)^2\right\rangle_{g,\Psi=0}=\left\langle\left(e^a_{\cV_1\cV_2}\right)^2\right\rangle_{g,\Psi=0},
\quad
\delta e^a_{\cV_1\cV_2}\equiv e^a_{\cV_1\cV_2}-\left\langle e^a_{\cV_1\cV_2}\right\rangle_{g,\Psi=0}.
\label{small40}
\end{gather}

Let's formulate the first conclusion which follows from  (\ref{small40}):

(i) The relations (\ref{small40}) show that the fluctuation values of tetrads $e_{{\cV}_1{\cV}_2}$ are commensurable with theirs most probable values
$\sqrt{\left\langle(e^a_{\cV_1\cV_2})^2\right\rangle_{g,\Psi=0}}$; the fluctuations are symmetric relative to the
inversion
$e_{{\cV}_1{\cV}_2}\longrightarrow -e_{{\cV}_1{\cV}_2}$ in pure gravity theory $\mA_g$ (\ref{Latt_Action_Grav}).

It is natural to expect that the first statement in  property (i) will also be preserved if fermions are included in the theory.
The obtained result justify the choice of model for propagation of irregular (doubled in  Wilson sense) fermion
quanta \footnote{By definition, the irregular quanta are that quanta which jump with a transition from
vertex to an adjacent vertex (see \cite{vergeles2015wilson}).}: the propagation of irregular quanta
on irregular "breathing" lattice is similar to the Markov process of a random
walks. So it turns out that the propagator of irregular modes on irregular lattice
decreases very quickly (exponentially): the doubled irregular modes
are "bad" quasiparticles \cite{vergeles2015wilson}.

It follows from (\ref{small30})  that in the theory of pure gravity, extended space-time does not exist.
Indeed, let's consider a successive  chain formed by  1-simplexes $\{a_{{\cV}_1}a_{{\cV}_2},\,\,
a_{{\cV}_2}a_{{\cV}_3},..., \,\,a_{{\cV}_{n-1}}a_{{\cV}_n}\}$. Then the distance between
the vertices $a_{{\cV}_1}$ and $a_{{\cV}_n}$ can be defined as follows:
\begin{gather}
\hat{L}_{12}=\hat{e}_{\cV_1\cV_2}+\Omega_{\cV_1\cV_2}\hat{e}_{\cV_2\cV_3}\Omega_{\cV_2\cV_1}+\ldots
+\Omega_{\cV_1\cV_2}\ldots\Omega_{\cV_{n-2}\cV_{n-1}}\hat{e}_{\cV_{n-1}\cV_n}
\Omega_{\cV_{n-1}\cV_{n-2}}\ldots\Omega_{\cV_2\cV_1}.
\label{small60}
\end{gather}
The quantity (\ref{small60}) depends not only  on the  vertices $a_{{\cV}_1}$ and $a_{{\cV}_n}$, but also on the chain
connecting  the  vertices and the method of parallel  shift of elementary vectors $\hat{e}_{(\cV_i\cV_j)}$ to the vertex $a_{\cV_1}$.  Due to (\ref{small32}) we have
\begin{gather}
\langle\hat{L}_{12}\rangle_{g,\Psi=0}=-\langle\hat{L}_{12}\rangle_{g,\Psi=0}=0.
\label{small70}
\end{gather}

In the gravity theory coupled with fermions with the action (\ref{Latt_Action})
the fluctuations symmetry relative to the
inversion $e_{{\cV}_1{\cV}_2}\longrightarrow -e_{{\cV}_1{\cV}_2}$   is broken.
Indeed, the fermion part of action (\ref{Latt_Action_Ferm}) is proportional to the third  power of the
variables $\{e\}$.  The considered system is modelled by the partition function $f(x)=A\exp\left((ia/2)x^2-
(i\lambda/3)x^3\right)$.
Then for $a\gg\lambda^{2/3}$:
\begin{gather}
\langle x\rangle\sim \lambda.
\label{small81}
\end{gather}
In our theory, the role of the constant $\lambda$ is played by the operator $\hat{\Theta}_{{\cV}_1{\cV}_2}$.

Let's consider the quantity
\begin{gather}
f_{({\cV}_1{\cV}_2)}(\varkappa)=\tr\hat{e}_{{\cV}_1{\cV}_2}\left\{\hat{e}_{{\cV}_1{\cV}_2}-
\varkappa_{({\cV}_1{\cV}_2)}\hat{\Theta}_{{\cV}_1{\cV}_2}\right\}.
\label{small82}
\end{gather}
Here $\varkappa$ is a numerical lattice function defined on 1-simplices. Eqs. (\ref{Gauge_Trans}), (\ref{Teta_Gauge_Trans}) and (\ref{PT_transform}), (\ref{PT_trans_Dir_Ac})  show that the quantity
$f_{({\cV}_1{\cV}_2)}(\varkappa)$ is gauge and PT invariant, and it follows from Eqs. (\ref{Variables_Grav}) and (\ref{Dir_Bil_Form_Trans})
that $f_{({\cV}_1{\cV}_2)}(\varkappa)=f_{({\cV}_2{\cV}_1)}(\varkappa)$.
Introduce the notation
\begin{gather}
\xi_{({\cV}_1{\cV}_2)}\equiv\Big\langle\tr(\hat{e}_{({\cV}_1{\cV}_2)})^2 \Big\rangle,
\quad \zeta_{({\cV}_1{\cV}_2)}\equiv\Big\langle\tr\hat{e}_{({\cV}_1{\cV}_2)}\hat{\Theta}_{({\cV}_1{\cV}_2)} \Big\rangle,
\label{Notations}
\end{gather}
Here the averaging $\langle\ldots\rangle$ means averaging according to (\ref{Partition_Function}) or (\ref{MeanField})  in case of $J\longrightarrow0$.
It can be argued that, by analogy with (\ref{small81}), the number $\zeta\neq0$ also. Indeed, the quantity
$\zeta$ is proportional to the vacuum average of the fermionic contribution to the energy-momentum tensor.
However, even in flat space-time, the vacuum average of the matter energy-momentum tensor   has the form
$\langle T^{ab}\rangle=\const\cdot\delta^{ab}$, where $\const\sim\l_P^{-4}$.
Thus we have:
\begin{gather}
\Big\langle f_{({\cV}_1{\cV}_2)}(\varkappa^{(0)})\Big\rangle=\xi_{({\cV}_1{\cV}_2)}-
\zeta_{({\cV}_1{\cV}_2)}\varkappa_{({\cV}_1{\cV}_2)}^{(0)}=0 \quad \mbox{for} \quad
\varkappa^{(0)}_{({\cV}_1{\cV}_2)}=\xi_{({\cV}_1{\cV}_2)}/\zeta_{({\cV}_1{\cV}_2)}.
\label{varkappa}
\end{gather}

Equality (\ref{varkappa}) leads us to the following assumption:
\begin{gather}
\frac{\delta}{\delta J^{(e)a}_{({\cV}_1{\cV}_2)}}\ln\mU\{J\}\bigg|_{J\longrightarrow0}=
\left\langle e_{{\cV}_1{\cV}_2}^a\right\rangle=\varkappa^{(0)}_{({\cV}_1{\cV}_2)}\left\langle \Theta^a_{{\cV}_1{\cV}_2}\right\rangle
\neq0
\label{Lat_tetrad_Aver}
\end{gather}
for some $\varkappa^{(0)}_{({\cV}_1{\cV}_2)}$.
 The relation (\ref{Lat_tetrad_Aver}) is also justified by the fact that the values under the average transform  identically under the transformations (\ref{Variables_Grav}) and
(\ref{Dir_Bil_Form_Trans}), (\ref{Gauge_Trans}) and (\ref{Teta_Gauge_Trans}),  (\ref{PT_transform}) and (\ref{PT_trans_Dir_Ac}).

In the long-wavelength limit, equality (\ref{Lat_tetrad_Aver}) becomes independent of the details of the lattice, but depends on the points of space-time:
\begin{gather}
\langle0|e^a_{\mu}(x)|0\rangle=\varkappa(x)\langle0|\Theta^a_{\mu}(x)|0\rangle.
\label{Long_Wav_Mean}
\end{gather}
The rationale for the latter equality is still based on the fact that both sides of the equality transform in the same way under the action of all the symmetries of action described here.
To evaluate the scalar function $\varkappa(x)$
we use the assumption that the quantum correlations between fermionic and gravitational degrees of freedom are very small. The reason for this is the fact that the theory of gravity itself is a non-renormalizable theory.
In nonrenormalizable theories, quantum fluctuations grow rapidly at short-wavelength
regions, but become insignificant in the long-wave region. This statement applies to gravitational but not fermionic variables. Therefore, the long-wavelength gravitational variables  can be considered "frozen" under vacuum averaging, that is, $\langle e^a_{\mu}\rangle=e^a_{\mu}$ and
$\langle\omega^{ab}_{\mu}\rangle=\omega^{ab}_{\mu}$. Thus, we obtain a chain of equalities for the vacuum mean of an invariant 4-volume element:
\begin{gather}
\langle 0|\d^{(4)}V|0\rangle
=\varepsilon_{abcd}e^a\wedge e^b\wedge e^c\wedge e^d=
\varkappa\varepsilon_{abcd}\langle 0|\Theta^a|0\rangle\wedge e^b\wedge e^c\wedge e^d
\sim\varkappa\langle 0|{\mL}_{\Psi}|0\rangle \d^{(4)}V.
\label{Inv_Vol_Mean}
\end{gather}
Here $\big(\mL_{\Psi}\d^{(4)}V \big)$ is the Dirac 4-form under the integral in (\ref{Long_Wav_Dir_Act}).
Since the quantity (\ref{Inv_Vol_Mean})  is  invariant relative to general coordinate transformations, we can estimate the mean $\langle0|\mL_{\Psi}|0\rangle\d^{(4)}V$ in normal Riemann coordinates, the center of which can be any point  $p$ of space-time. In normal Riemann coordinates $y^{\mu}$ we have $e^a_{\mu}|_p=\delta^a_{\mu}$, $\omega^{ab}_{\ mu}|_p=0$. Therefore, near the point $p$, the required value is estimated in the same way as in flat space:
\begin{gather}
\langle0|\mL_{\Psi}|0\rangle\d^{(4)}V\sim (2\pi\hbar)^4\left(\frac{\mN_{\Psi}}{\mV_{\bf x}}\right)^{4/3}.
\label{Ferm_Action_Mean}
\end{gather}
Here $\mV_{\bf x}$ denotes a purely spatial volume, and $\mN_{\Psi}$ is the number of negative-frequency Dirac modes in this volume. Comparing (\ref{Inv_Vol_Mean}) and (\ref{Ferm_Action_Mean}), we come to the estimate
\begin{gather}
 \varkappa\sim(2\pi\hbar)^{-4}\left(\frac{\mV_{\bf x}}{\mN_{\Psi}}\right)^{4/3}.
 \label{varkappa}
\end{gather}

In the case of Friedmann's models or similar ones (which is realized in the Universe), space is homogeneous and isotropic. Then, according to (\ref{varkappa}), since $\mN_{\Psi}=\const$, the function $\varkappa$ does not depend on spatial coordinates and is proportional to the fourth power of the scale factor of the Universe.

Now we can formulate the second conclusion:

(ii) In the lattice theory of pure gravity under consideration the macroscopic space-time cannot exist.
However, in the presence of fermions, the possibility of the “birth” of macroscopic space-time appears.

\section{Conclusion}

Let's summarize the obtained conclusions.

(i) The fluctuation values of tetrads $e_{\cV_1\cV_2}$ are commensurable with theirs most probable values
$\sqrt{\left\langle(e^a_{\cV_1\cV_2})^2\right\rangle}$; the fluctuations are symmetric relative to the
inversion $e_{{\cV}_1{\cV}_2}\longrightarrow -e_{{\cV}_1{\cV}_2}$ in pure gravity theory (\ref{Latt_Action_Grav}).
But in the lattice theory of gravity coupled with fermions, the latter property is violated according to (\ref{Lat_tetrad_Aver}).

(ii) In the lattice theory of pure gravity under consideration the macroscopic space-time cannot exist.
However, in the presence of fermions, the possibility of the “birth” of macroscopic space-time appears.

(iii) Despite the fact that the full action (\ref{Latt_Action}) is  PT-invariant,  the experimenter will record a violation of PT-parity. So,
PT-invariance is violated in the considered lattice theory of gravity coupled with fermions.
The described PT  symmetry breaking can be considered as a spontaneous symmetry breaking.

Since the transformation (\ref{PT_transform}) or (\ref{PT_trans_Cont_Minkovski})  rearranges mutually  fermionic particles and antiparticles, the described vacuum structure breaks the symmetry between fermions and antifermions.
Thus, a debatable question arises: can the inclusion of gravity help to clarify the mechanism of CP violation
beyond the Standard Model, which fails to explain the observed baryon asymmetry of the Universe
\cite{sakharov1991violation,bernreuther2002cp,chupp2019electric}?

\begin{acknowledgments}

I am grateful to G. Volovik for stimulating attention to this work.
This work was carried out as a part of the State Program 0033-2019-0005.

\end{acknowledgments}


\begin{thebibliography}{26}
\expandafter\ifx\csname natexlab\endcsname\relax\def\natexlab#1{#1}\fi
\expandafter\ifx\csname bibnamefont\endcsname\relax
  \def\bibnamefont#1{#1}\fi
\expandafter\ifx\csname bibfnamefont\endcsname\relax
  \def\bibfnamefont#1{#1}\fi
\expandafter\ifx\csname citenamefont\endcsname\relax
  \def\citenamefont#1{#1}\fi
\expandafter\ifx\csname url\endcsname\relax
  \def\url#1{\texttt{#1}}\fi
\expandafter\ifx\csname urlprefix\endcsname\relax\def\urlprefix{URL }\fi
\providecommand{\bibinfo}[2]{#2}
\providecommand{\eprint}[2][]{\url{#2}}

\bibitem[{\citenamefont{Vergeles}(2015)}]{vergeles2015wilson}
\bibinfo{author}{\bibfnamefont{S.}~\bibnamefont{Vergeles}},
  \bibinfo{journal}{Physical Review D} \textbf{\bibinfo{volume}{92}},
  \bibinfo{pages}{025053} (\bibinfo{year}{2015}).

\bibitem[{\citenamefont{Vergeles}(2016)}]{vergeles2016instanton}
\bibinfo{author}{\bibfnamefont{S.~N.} \bibnamefont{Vergeles}},
  \bibinfo{journal}{JETP letters} \textbf{\bibinfo{volume}{104}},
  \bibinfo{pages}{494} (\bibinfo{year}{2016}).

\bibitem[{\citenamefont{Vergeles}(2006)}]{vergeles2006one}
\bibinfo{author}{\bibfnamefont{S.}~\bibnamefont{Vergeles}},
  \bibinfo{journal}{Nuclear Physics B} \textbf{\bibinfo{volume}{735}},
  \bibinfo{pages}{172} (\bibinfo{year}{2006}).

\bibitem[{\citenamefont{Vergeles}(2008)}]{vergeles2008doubling}
\bibinfo{author}{\bibfnamefont{S.}~\bibnamefont{Vergeles}},
  \bibinfo{journal}{Journal of Experimental and Theoretical Physics}
  \textbf{\bibinfo{volume}{106}}, \bibinfo{pages}{46} (\bibinfo{year}{2008}).

\bibitem[{\citenamefont{Vergeles}(2017{\natexlab{a}})}]{vergeles2017note}
\bibinfo{author}{\bibfnamefont{S.}~\bibnamefont{Vergeles}},
  \bibinfo{journal}{Journal of High Energy Physics}
  \textbf{\bibinfo{volume}{2017}}, \bibinfo{pages}{44}
  (\bibinfo{year}{2017}{\natexlab{a}}).

\bibitem[{\citenamefont{Vergeles}(2017{\natexlab{b}})}]{vergeles2017fermion}
\bibinfo{author}{\bibfnamefont{S.}~\bibnamefont{Vergeles}},
  \bibinfo{journal}{Physical Review D} \textbf{\bibinfo{volume}{96}},
  \bibinfo{pages}{054512} (\bibinfo{year}{2017}{\natexlab{b}}).

\bibitem[{\citenamefont{Regge}(1961)}]{regge1961general}
\bibinfo{author}{\bibfnamefont{T.}~\bibnamefont{Regge}}, \bibinfo{journal}{Il
  Nuovo Cimento (1955-1965)} \textbf{\bibinfo{volume}{19}},
  \bibinfo{pages}{558} (\bibinfo{year}{1961}).

\bibitem[{\citenamefont{Diakonov}(2011)}]{diakonov2011towards}
\bibinfo{author}{\bibfnamefont{D.}~\bibnamefont{Diakonov}},
  \bibinfo{journal}{arXiv preprint arXiv:1109.0091}  (\bibinfo{year}{2011}).

\bibitem[{\citenamefont{Vladimirov and Diakonov}(2012)}]{vladimirov2012phase}
\bibinfo{author}{\bibfnamefont{A.~A.} \bibnamefont{Vladimirov}}
  \bibnamefont{and} \bibinfo{author}{\bibfnamefont{D.}~\bibnamefont{Diakonov}},
  \bibinfo{journal}{Physical Review D} \textbf{\bibinfo{volume}{86}},
  \bibinfo{pages}{104019} (\bibinfo{year}{2012}).

\bibitem[{\citenamefont{Hamber}(2009)}]{hamber2009quantum}
\bibinfo{author}{\bibfnamefont{H.~W.} \bibnamefont{Hamber}},
  \bibinfo{journal}{General Relativity and Gravitation}
  \textbf{\bibinfo{volume}{41}}, \bibinfo{pages}{817} (\bibinfo{year}{2009}).

\bibitem[{\citenamefont{Bianchi et~al.}(2013)\citenamefont{Bianchi, Han,
  Rovelli, Wieland, Magliaro, and Perini}}]{bianchi2013spinfoam}
\bibinfo{author}{\bibfnamefont{E.}~\bibnamefont{Bianchi}},
  \bibinfo{author}{\bibfnamefont{M.}~\bibnamefont{Han}},
  \bibinfo{author}{\bibfnamefont{C.}~\bibnamefont{Rovelli}},
  \bibinfo{author}{\bibfnamefont{W.}~\bibnamefont{Wieland}},
  \bibinfo{author}{\bibfnamefont{E.}~\bibnamefont{Magliaro}}, \bibnamefont{and}
  \bibinfo{author}{\bibfnamefont{C.}~\bibnamefont{Perini}},
  \bibinfo{journal}{Classical and Quantum Gravity}
  \textbf{\bibinfo{volume}{30}}, \bibinfo{pages}{235023}
  (\bibinfo{year}{2013}).

\bibitem[{\citenamefont{Engle et~al.}(2008)\citenamefont{Engle, Livine,
  Pereira, and Rovelli}}]{engle2008lqg}
\bibinfo{author}{\bibfnamefont{J.}~\bibnamefont{Engle}},
  \bibinfo{author}{\bibfnamefont{E.}~\bibnamefont{Livine}},
  \bibinfo{author}{\bibfnamefont{R.}~\bibnamefont{Pereira}}, \bibnamefont{and}
  \bibinfo{author}{\bibfnamefont{C.}~\bibnamefont{Rovelli}},
  \bibinfo{journal}{Nuclear Physics B} \textbf{\bibinfo{volume}{799}},
  \bibinfo{pages}{136} (\bibinfo{year}{2008}).

\bibitem[{\citenamefont{Perez}(2013)}]{perez2013spin}
\bibinfo{author}{\bibfnamefont{A.}~\bibnamefont{Perez}},
  \bibinfo{journal}{Living Reviews in Relativity}
  \textbf{\bibinfo{volume}{16}}, \bibinfo{pages}{1} (\bibinfo{year}{2013}).

\bibitem[{\citenamefont{Volovik}(2003)}]{volovik2003universe}
\bibinfo{author}{\bibfnamefont{G.~E.} \bibnamefont{Volovik}},
  \emph{\bibinfo{title}{The universe in a helium droplet}}, vol.
  \bibinfo{volume}{117} (\bibinfo{publisher}{Oxford University Press on
  Demand}, \bibinfo{year}{2003}).

\bibitem[{\citenamefont{Nissinen and Volovik}(2018)}]{nissinen2018dimensional}
\bibinfo{author}{\bibfnamefont{J.}~\bibnamefont{Nissinen}} \bibnamefont{and}
  \bibinfo{author}{\bibfnamefont{G.}~\bibnamefont{Volovik}},
  \bibinfo{journal}{Physical Review D} \textbf{\bibinfo{volume}{97}},
  \bibinfo{pages}{025018} (\bibinfo{year}{2018}).

\bibitem[{\citenamefont{Volovik}(1990)}]{volovik1990superfluid}
\bibinfo{author}{\bibfnamefont{G.}~\bibnamefont{Volovik}},
  \bibinfo{journal}{Physica B: Condensed Matter}
  \textbf{\bibinfo{volume}{162}}, \bibinfo{pages}{222} (\bibinfo{year}{1990}).

\bibitem[{\citenamefont{Wetterich}(2012)}]{wetterich2012universality}
\bibinfo{author}{\bibfnamefont{C.}~\bibnamefont{Wetterich}},
  \bibinfo{journal}{Physics Letters B} \textbf{\bibinfo{volume}{712}},
  \bibinfo{pages}{126} (\bibinfo{year}{2012}).

\bibitem[{\citenamefont{Sexty and Wetterich}(2013)}]{sexty2013emergent}
\bibinfo{author}{\bibfnamefont{D.}~\bibnamefont{Sexty}} \bibnamefont{and}
  \bibinfo{author}{\bibfnamefont{C.}~\bibnamefont{Wetterich}},
  \bibinfo{journal}{Nuclear Physics B} \textbf{\bibinfo{volume}{867}},
  \bibinfo{pages}{290} (\bibinfo{year}{2013}).

\bibitem[{\citenamefont{Salomaa and Volovik}(1987)}]{salomaa1987quantized}
\bibinfo{author}{\bibfnamefont{M.}~\bibnamefont{Salomaa}} \bibnamefont{and}
  \bibinfo{author}{\bibfnamefont{G.}~\bibnamefont{Volovik}},
  \bibinfo{journal}{Reviews of modern physics} \textbf{\bibinfo{volume}{59}},
  \bibinfo{pages}{533} (\bibinfo{year}{1987}).

\bibitem[{\citenamefont{Salomaa and Volovik}(1988)}]{salomaa1988cosmiclike}
\bibinfo{author}{\bibfnamefont{M.}~\bibnamefont{Salomaa}} \bibnamefont{and}
  \bibinfo{author}{\bibfnamefont{G.}~\bibnamefont{Volovik}},
  \bibinfo{journal}{Physical Review B} \textbf{\bibinfo{volume}{37}},
  \bibinfo{pages}{9298} (\bibinfo{year}{1988}).

\bibitem[{\citenamefont{Nissinen and Volovik}(2017)}]{nissinen2017effective}
\bibinfo{author}{\bibfnamefont{J.}~\bibnamefont{Nissinen}} \bibnamefont{and}
  \bibinfo{author}{\bibfnamefont{G.~E.} \bibnamefont{Volovik}},
  \bibinfo{journal}{JETP Letters} \textbf{\bibinfo{volume}{106}},
  \bibinfo{pages}{234} (\bibinfo{year}{2017}).

\bibitem[{\citenamefont{Sakharov}(1991)}]{sakharov1991violation}
\bibinfo{author}{\bibfnamefont{A.}~\bibnamefont{Sakharov}},
  \bibinfo{journal}{Sov. Phys. Usp} \textbf{\bibinfo{volume}{34}},
  \bibinfo{pages}{392} (\bibinfo{year}{1991}).

\bibitem[{\citenamefont{Bernreuther}(2002)}]{bernreuther2002cp}
\bibinfo{author}{\bibfnamefont{W.}~\bibnamefont{Bernreuther}}, in
  \emph{\bibinfo{booktitle}{CP Violation in Particle, Nuclear and
  Astrophysics}} (\bibinfo{publisher}{Springer}, \bibinfo{year}{2002}), pp.
  \bibinfo{pages}{237--293}.

\bibitem[{\citenamefont{Chupp et~al.}(2019)\citenamefont{Chupp, Fierlinger,
  Ramsey-Musolf, and Singh}}]{chupp2019electric}
\bibinfo{author}{\bibfnamefont{T.}~\bibnamefont{Chupp}},
  \bibinfo{author}{\bibfnamefont{P.}~\bibnamefont{Fierlinger}},
  \bibinfo{author}{\bibfnamefont{M.}~\bibnamefont{Ramsey-Musolf}},
  \bibnamefont{and} \bibinfo{author}{\bibfnamefont{J.}~\bibnamefont{Singh}},
  \bibinfo{journal}{Reviews of Modern Physics} \textbf{\bibinfo{volume}{91}},
  \bibinfo{pages}{015001} (\bibinfo{year}{2019}).

\bibitem[{\citenamefont{Boyle et~al.}(2018)\citenamefont{Boyle, Finn, and
  Turok}}]{boyle2018c}
\bibinfo{author}{\bibfnamefont{L.}~\bibnamefont{Boyle}},
  \bibinfo{author}{\bibfnamefont{K.}~\bibnamefont{Finn}}, \bibnamefont{and}
  \bibinfo{author}{\bibfnamefont{N.}~\bibnamefont{Turok}},
  \bibinfo{journal}{Physical review letters} \textbf{\bibinfo{volume}{121}},
  \bibinfo{pages}{251301} (\bibinfo{year}{2018}).

\bibitem[{\citenamefont{t~Hooft and Veltman}(1974)}]{t1974one}
\bibinfo{author}{\bibfnamefont{G.}~\bibnamefont{t~Hooft}} \bibnamefont{and}
  \bibinfo{author}{\bibfnamefont{M.}~\bibnamefont{Veltman}}, in
  \emph{\bibinfo{booktitle}{Annales de l'IHP Physique th{\'e}orique}}
  (\bibinfo{year}{1974}), vol.~\bibinfo{volume}{20}, pp.
  \bibinfo{pages}{69--94}.

\end{thebibliography}

\end{document}